\documentclass[twocolumn,prd,showpacs,10pt,nofootinbib]{revtex4}
\usepackage{amsmath,amsfonts,amssymb,bm,epsfig}

\begin{document}


\title[Short title for running header]{Cosmic Acceleration from Elementary Interactions}
\author{R. Aldrovandi$^1$\footnote{Electronic address: ra@ift.unesp.br},
R. R. Cuzinatto$^1$\footnote{Electronic address:
rodrigo@ift.unesp.br} and L. G. Medeiros$^1$\footnote{Electronic
address: leo@ift.unesp.br}}

\address{$^1$ Instituto de F\'{\i}sica Te\'{o}rica, Universidade Estadual
Paulista. \\ Rua Pamplona 145, CEP 01405-000, S\~{a}o Paulo, SP,
Brazil}


\begin{abstract}
 It is possible to generate an accelerated period of expansion
from reasonable potentials acting between the universe particle
constituents. The pressure of primordial nucleons interacting via a
simple nuclear potential is obtained via  Mayer's cluster expansion
technique.  The attractive part of the potential engenders a
negative pressure and may therefore be responsible for the cosmic
acceleration.
\end{abstract}


\maketitle


\section{Introduction}

Cosmic primordial acceleration is usually derived from a scalar field.
The idea of using a scalar field arose in the context of
GUT's~\cite{Guth}, at first with the Higgs boson field in mind.  This
attribution has been dropped in the long run, as the Higgs field did
not reproduce all necessary features required by the inflationary
period~\cite{Liddle}.  As a consequence the inflaton -- an \emph{ad
hoc} scalar field leading to the desired properties -- was introduced.  This
arbitrary field, however, has no simple interpretation in terms of
fundamental physical phenomena~---~which justifies the search for
alternative explanations.

 We intend here to show that primeval acceleration may come from the
 strong interactions in the constituents' equation of state.  Such short
 range interactions are taken into account via the well--known and
 systematic method of cluster expansions~\cite{pathria}.

An application to cosmology of a toy-model equation of state (EOS) for
systems with interaction has been made in \cite{ruben}, where the
elementary constituents of the universe were taken to be hard-spheres.
A recent paper \cite{bascos} has proposed an explanation for the
inflationary period through a short range interaction, just what we
shall do.  Nevertheless, the modification on the equation for the
pressure was not obtained through the Mayer method here employed,
which brings to light a time scale determined by the temperature and
leads to a natural passage from the accelerated period to the
decelerated one.

In section \ref{sec-PreNucl} we synthesize the main characteristics of
our model, specifying the relevant particles of the matter content.
Section \ref {sec-EOS} deals with the construction of the equation of
state for the interacting media using Mayer's approach of virial
cluster expansion.  The nuclear interaction is modeled by the
effective potentials presented in section \ref{sec-graphics}.  The
results for our simplistic model point to a
period of accelerated expansion and are discussed in section \ref%
{sec-conclusion}. \ \


\section{On the pre-nucleosynthesis period\label{sec-PreNucl}}

The pre-nucleosynthesis period is the stage immediately preceding the
cosmological  formation of the light elements  (deuteron, He$^{+2}$, etc).
This formation only started when the
universe mean energy attained values around the deuteron binding
energy $(E_{B}\simeq 2.23MeV)$. So, we roughly characterize  this period by the red-shift  $z \gtrsim 2 \times 10^{12}$ or,  in
terms of energy,  $kT \gtrsim  4MeV$.

Radiation dominates the universe content in this phase. Photons generate a panoply of particles via  pair production,  with populations dependent on  the energy of the thermalized system~\cite{Kolb,Nar}. In particular, at hundreds of $MeV$ nucleons produced by this process are in chemical equilibrium
\cite{Zel}. At these energies the numerical density  $n_{N}$ of nucleons
deriving from the radiation  is much superior to that ($n_{b}$) of the
present-day nucleons, which can be consequently  neglected.

A realistic scenario takes into account, besides the nucleons, light
particles appearing even before the nucleons in the pair production
process. Examples are pions, eletron-positron pairs, muons. For
simplicity, we will adopt as relevant content photons, nucleons and
anti-nucleons (protons and anti-protons, neutrons and
anti-neutrons). We will neglect eletrons and positrons, pions,
neutrinos, muons and anti-muons.

The presence of charged particles would require the consideration of the
electromagnetic interaction among the constituents as a source for
the gravitational field. However, $n_{N} >> n_{b}$ implies
symmetry between matter and anti-matter. We shall suppose overall
 Debye screening: in large enough scales, the system will be electrically neutral. Once the electromagnetic interaction is neglected, we are left with
the strong and weak interactions. Obviously, the weak nuclear interaction is much less intense
compared with the strong one: the strength of the former is
$10^{-13} $ orders of magnitude of the last \cite{griffiths}. This
justifies the approximation we will adopt: our equation of state for
the pre-nucleosynthesis period considers only the strong nuclear
interaction. In short, our model ($\gamma +N \bar{N}$) is constituted
by photons plus interacting nucleons treated as classical
non-relativistic particles obeying Boltzmann statistics.


\section{EOS of a system under interaction \label{sec-EOS}}

With the hypothesis introduced above, the equation of state of the
system
composed by photons ($\gamma $) and nucleons/antinucleons ($N/\bar{N}$) will have the form:
\begin{equation}
p(kT, n_{N})=p_{\gamma }(kT)+p_{N}(kT,n_{N}),  \label{partial pressure}
\end{equation}%
where $kT$ is the energy and $n_{N}$ is the nucleon$+$antinucleon
numerical density.

We shall approach the calculation of the strong interaction between
nucleons by Mayer's method~\cite{pathria}. In this formalism, the
pressure of a classic gas of interacting particles
is calculated through the virial expansion:%
\begin{equation}
p_{N}(kT,n_{N})=n_{N}kT\sum\limits_{l=1}^{\infty }a_{l}(T) \left(
\frac{n_{N}\lambda ^{3}}{g} \right)^{l-1},  \label{virial expansion}
\end{equation}
$a_{l}(T)$ being the virial coefficients, \textit{viz.}
\begin{subequations}
\begin{eqnarray}
a_{1} &=&1;~\ \ a_{2}=-\frac{2\pi }{\lambda ^{3}}\int\limits_{0}^{\infty
}(e^{-u(r)/kT}-1)r^{2}dr;   \\
a_{3} &=&\frac{-1}{3\lambda ^{6}}\iint\limits_{0}^{\infty
}f_{12}f_{13}f_{23}d^{3}r_{12}d^{3}r_{13};~  \label{virial coefficients}
\end{eqnarray}
\end{subequations}
and so on, given in terms of the mean thermal wavelength $\lambda =$ $h/%
\sqrt{2\pi m_N kT}$ of the nucleons, and of the two-particle Mayer function $%
f_{ij}=e^{-\beta u_{ij}}-1$ which  depends on $\beta =1/kT$ and on
the inter-particle potential $u_{ij}=u(|\vec{r}_{i}-\vec{r}_{j}|)$.
The second virial coefficient $a_{2}$ in (\ref{virial coefficients})
reflects the interaction of the particles two by two (the
2-cluster); and $a_{3}$\ accounts for the interaction among three
particles. $g$\ is the number of possible values for the internal
degrees of freedom. Four species are to be taken into account
(protons, anti-proton, neutron and anti-nucleon). Besides, our
species have spin $\frac{1}{2}$: $g=4\times2=8$.

Under our assumptions, only short-range nuclear forces will be at work.
It is consequently reasonable to admit that interaction occurs only
between few particles, and  that $p_{N}$ as given by (\ref{virial
expansion}) can be conveniently approximated by its first three
terms. This hypothesis is  reasonable as long as the density is not too high. Substituting (\ref{virial expansion}) in the equality (\ref{partial pressure}) leads to the EOS
\begin{gather}
p(n_{N},kT)=\frac{\pi ^{2}}{45\hbar ^{3}c^{3}}(kT)^{4}+n_{N}kT+  \notag \\
 + n_{N}kT\left[ a_{2}(kT)\left( \frac{n_{N} \lambda^3}{g} \right)
+ a_3(kT) \left( \frac{n_N \lambda^3}{g} \right)^2 \right] .
\label{EOS}
\end{gather}%
The first term, $p_{\gamma }(kT)$, comes  from the
black-body thermodynamics.

The natural time parameter along the thermalized period  predating recombination is just $kT$. To examine  the history of  Eq.(\ref{EOS}), it would be convenient  to express it in terms of this parameter. This requires the determination of $n_{N}=n_{N}(kT)$. Supposing  equality between the nucleon and antinucleon concentrations and masses, the
numerical density will be
\begin{eqnarray}
n_{N}(kT) &=& \frac{g}{\lambda ^{3}}\left[ e^{\frac{\mu
_{N}}{kT}}-2a_{2}e^{2\frac{\mu _{N}}{kT}}+\right.  \notag \\
&&\left. +3\left( \frac{4a_{2}^{~2}-a_{3}}{2}\right) e^{3\frac{\mu _{N}}{kT}}%
\right] ,  \label{numerical density}
\end{eqnarray}%
where $\mu_{N}$\ is the nucleon chemical potential.

The system in equilibrium must assure the formation of the nucleons,
\textit{i.e.}, $\mu _{N} = - m_ N c^2$.


\section{Graphics of $p(kT)$ for  Phenomenological Potentials\label%
{sec-graphics}}

\begin{figure}[h]
 \begin{center}
 \centerline{\epsfig{file=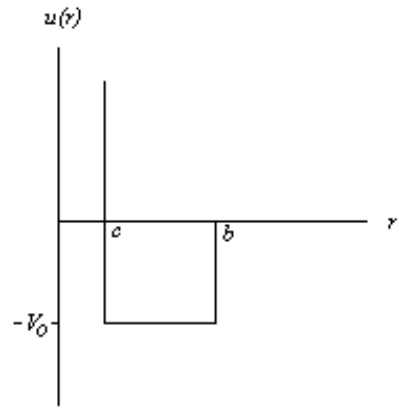,height=6.0cm}}
 \caption{{\it Scheme of the supposed strong interaction potential.}}
 \label{fig-pot}
 \end{center}
\end{figure}

In a simplistic manner, we describe the strong interaction by  a
nuclear potential of the square-well+hard-core type shown in Figure
\ref{fig-pot}.
The choice of  this potential is justified by two facts: (i) it shows a
behavior qualitatively similar to some successful phenomenological nuclear
potentials, \textit{e.g.}, those discussed in~\cite{Reid,Hammada}; (ii)
it makes it possible to calculate analytically the corrections to the ideal case.

The potential parameters are determined from the deuteron binding
energy, from its mean square radius~---~which constrain the well width
 $ \left( b-c\right) =1.3$~fm  and its depth\ $V_{0}=75.6$~MeV~---~and from
nucleons scattering data at high energies~---~which determinate the hard core
extension $c=0.4$~fm (\textit{cf.} reference~\cite{enge}).

\begin{figure}[h]
 \begin{center}
 \centerline{\epsfig{file=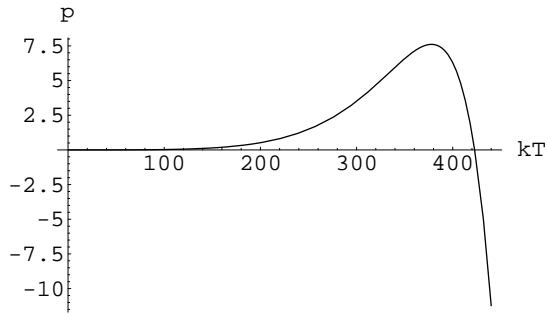,height=4.5cm}}
 \caption{{\it  Pressure of the interacting
nucleons plus photons as a function of the energy:
given in units of  $MeV/\lambda _{c}^{3}$, where $\lambda _{c}=\hbar
c/m_{N}c^{2}$, and the energy is given in $MeV$.}}
 \label{fig-pressure}
 \end{center}
\end{figure}

Function $p(kT)$ is obtained by substituting (\ref{numerical
density}) in (\ref{EOS}) and using the $a_{2}(kT)$ and $a_{3}(kT)$
explicit expressions. Its aspect is shown in Figure
\ref{fig-pressure}.
It exhibits an increasing behavior at low energies ($kT < 330$ MeV).
This is only natural,  since at these energies the nucleons density
is still too small to cause a relevant interaction effect. A sudden
change occurs, however,  at around $375$ MeV. The pressure starts
decreasing as the energy rises, becoming negative at around $420$
MeV. This surprising behavior is due to the action of the
interaction term $a_{2}(kT)$ which dominates in (\ref{EOS}). To be
more specific: it is the attractive part (negative sector) of the
potential in $a_{2}(kT)$\ that surpasses all the other terms.

It is worth noticing that for energies until $490$ $MeV$ the two by
two interaction term ($a_{2}(kT)n_{N}\lambda ^{3}$) between nucleons
is at least five times larger than the three by three interaction
term ($ a_{3}(kT)n_{N}^{2}\lambda ^{6}$). This is a point  in favor
of truncating  the series (\ref{virial expansion}) in the first
terms.


\section{Final Remarks\label{sec-conclusion}}

As the pressure, the energy density $\rho _{N}(kT,n_{N})$\ may be
expressed in terms of a series of type (\ref{virial expansion}). The
method is the same presented in section
\ref{sec-EOS}\ and the
result is:
\begin{gather}
\rho (kT,n_{N})=\frac{\pi ^{2}}{15\hbar ^{3}c^{3}}(kT)^{4}+n_{N}\left(
m_{N}c^{2}+\frac{3kT}{2}\right) +  \notag \\
+ \frac{3}{2} kT \frac{n_N^{~2} \lambda^{3}}{g} \left[ \left(
a_{2}-\frac{2T}{3}\frac{\partial a_{2}}{\partial T}\right) + \left(
a_{3}-\frac{T}{3} \frac{\partial a_{3}}{\partial T}\right) \frac{n_N
\lambda^3}{g} \right] , \label{energy density}
\end{gather}
$m_{N}$\ being, we recall, the nucleon mass.

\begin{figure}[h]
 \begin{center}
 \centerline{\epsfig{file=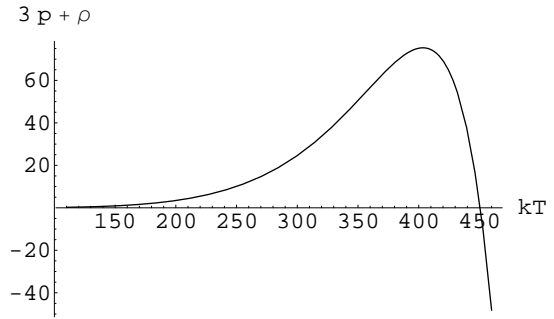,height=4.5cm}}
 \caption{{\it  Plot of $\rho +3p$
 as a function of $kT$ in $MeV$. In our units, $4\pi G/3c^{2}=1$. Recall  that  the cosmic time and  the
scale factor decrease as the temperature increases.}}
 \label{fig-rho_p}
 \end{center}
\end{figure}

With the pressure equation (\ref{EOS}) and the equality (\ref
{energy density}),  we can rewrite  the Friedmann equation
\begin{equation}
\frac{\ddot{a}}{a}=-\frac{4\pi G}{3c^{2}}(\rho +3p)  \label{Friedmann}
\end{equation}%
in terms of $kT$. The behavior of $\rho +3p$ is given in Figure
\ref{fig-rho_p}.

Combined with (\ref{Friedmann}), it shows  that positive
acceleration occurs
 from $450$ $MeV$ on. A detailed analysis of
 the Friedmann equations brings forth a division in the
thermal history of the model with interactions in four distinct
periods:

\begin{equation*}
\begin{tabular}{|c|c|}
\hline Energy ($MeV$) & Thermal history phase \\ \hline $0\lesssim
kT<100$ & decelerated expansion $\left( p=\frac{\rho }{3}\right) $
\\ \hline
$100\leq kT<450$ & transition phase $\left( -\frac{\rho }{3}\leq
p<\frac{\rho }{3}\right) $ \\ \hline $450\leq kT<470$ & accelerated
$\left( -\rho \leq p<-\frac{\rho }{3}\right) $
\\ \hline
$ kT \geq 470$ & ghostly accelerated $\left( p<-\rho \right) $ \\
\hline
\end{tabular}
\end{equation*}
TABLE:\textit{\ Thermal history phases of the model with
interaction. Notice how the model links naturally the accelerated
expansion period with a radiation type decelerated era.}

\bigskip

Further improvement would come from taking into account the
neglected particles. The nuclear interaction could also be
approached in a more sophisticated way, \textit{e.g.}, by using the
phenomenological potentials given in refs. \cite{Reid,Hammada},
introducing quantum corrections \emph{a la}\ Bethe-Uhlenbeck
\cite{pathria}, adding relativistic  corrections, etc.

The idea that an attractive potential between the constituents can led
to  accelerated expansion can be tentatively applied to the present day observed acceleration. The scenario is then that of   matter interacting through the gravitational potential. The latter being  attractive, the expected
global result would be a decreasing in the value of the pressure
towards negative values, eventually causing the acceleration. Around the
same period, but locally, the gravitational potential
would became sufficiently effective to engender the large structures
in the universe. This is consistent with the observational
data,  which indicate that both the accelerated expansion~\cite
{spernova} and the large scale structure formation~\cite{HST} have begun recently.


\begin{center}
\section*{ACKNOWLEDGMENTS}
\end{center}

R. R. C. and L. G. M. are grateful to Funda\c{c}\~{a}o de Amparo
\`{a} Pesquisa do Estado de S\~ao Paulo (FAPESP), Brazil,  for
support (grants 02/05763-8 and 02/10263-4 respectively) and R. A.
thanks to Conselho Nacional de Pesquisas (CNPq), Brazil.



\begin{thebibliography}{99}

\bibitem{Guth} A. H. Guth, \textit{Inflationary Universe: A Possible
Solution to the Horizon and Flatness Problem}, Phys. Rev. D \textbf{23}, 357
(1981).

\bibitem{Liddle} A. R. Liddle and D. H. Lyth, \textit{Cosmological Inflation
and Large-Scale Structure}, Cambridge University Press, 2000.

\bibitem{pathria} R.K. Pathria, \textit{Statistical Mechanics,} 1st. ed.,
Butterworth Heinemann, Oxford, 1972.

\bibitem{ruben} R. Aldrovandi, J. Gariel and G. Marcilhacy, \textit{On the
Pre-nucleosynthesis Cosmological Period}, {\em Revista Ci\^encias Exatas e Naturais}, {\bf 5} (2) 133-156 (2003)~---~gr-qc/0203079 (2002).

\bibitem{bascos} A D\'{\i}ez-Tejedor and A. Feinstein, \textit{Accelerating
Universes from Short-Range Interactions}, gr-qc/0505105 (2005), to appear in Phys. Lett. A.

\bibitem{Kolb} E. W. Kolb and M. S. Turner, \textit{The Early Universe},
Perseus Books, 1994.

\bibitem{Nar} J. V. Narlikar, \textit{Introduction to Cosmology,} 2nd ed.,
Cambridge University Press, Cambridge, 1993.

\bibitem{Zel} Ya. B. Zeldovich and I. D. Novikov,\textit{\ Relativistic
Astrophysics II: The Structure and Evolution of the Universe,} University of
Chicago Press, 1981.

\bibitem{enge} H. Enge, \textit{Introduction to Nuclear Physics,}
Addison-Wesley,1966.


\bibitem{griffiths} D. Griffiths, \textit{Introduction to Elementary Particles}, John Wiley, New york, 1987.

\bibitem{Reid} R. Reid, \textit{Local Phenomenological Nucleon-Nucleon
Potentials}, Annals of Physics \textbf{50}, 411, 1968.

\bibitem{Hammada} T. Hamada and I. D. Johnston, \textit{A Potential Model
Representation of Two-Nucleon Data Below 315 MeV}, Nuclear Physics \textbf{34%
}, 382, 1962.

\bibitem{spernova} S. Permutter et al.,\textit{\ Discovery of a Supernova
Explosion at Half the Age of the Universe, }Nature 392, 51 (1998).\textit{\ }

\bibitem{HST} W. L. Freedman et al., \textit{Final Results from the Hubble
Space Telescope Key Project to Measure the Hubble Constant}, Astrophysics
Journal 553, 47, 2001.
\end{thebibliography}
\end{document}